\documentclass[11pt]{article}

\usepackage{times}
\usepackage{graphicx}
\usepackage{multirow}
\usepackage{rotating}
\usepackage{amsbsy}
\usepackage{natbib}

\makeatletter
\def\maketitle{\par 
\begingroup
   \def\thefootnote{\fnsymbol{footnote}}
   \def\@makefnmark{\hbox to 0pt{$^{\@thefnmark}$\hss}} 
   \long\def\@makefntext##1{\parindent 1em\noindent \hbox to1.8em{\hss $\m@th ^{\@thefnmark}$}##1}
   \@maketitle \@thanks
\endgroup
\setcounter{footnote}{0}
\let\maketitle\relax \let\@maketitle\relax
\gdef\@thanks{}\gdef\@author{}\gdef\@title{}\let\thanks\relax}
\makeatother

\makeatletter
\def\@maketitle{\vbox{\hsize\textwidth
\linewidth\hsize \vskip 0.1in \toptitlebar \centering
{\LARGE\bf \@title\par}  \bottomtitlebar 
   \def\And{\end{tabular}\hfil\linebreak[0]\hfil
            \begin{tabular}[t]{c}\bf\rule{\z@}{24pt}\ignorespaces}%
   \def\AND{\end{tabular}\hfil\linebreak[4]\hfil
            \begin{tabular}[t]{c}\bf\rule{\z@}{24pt}\ignorespaces}%
   \def\LINEBREAK{\end{tabular}\linebreak[4]\begin{tabular}[t]{c}\bf\rule{\z@}{16pt}\ignorespaces}%
    \begin{tabular}[t]{c}\bf\rule{\z@}{24pt}\@author\end{tabular}%
\vskip 0.3in minus 0.1in}}
\makeatother

\renewenvironment{abstract}{\vskip.075in\centerline{\large\bf Abstract}\vspace{0.5ex}\begin{quote}}{\par\end{quote}\vskip 1ex}

\makeatletter
\def\section{\@startsection {section}{1}{\z@}{-2.0ex plus -0.5ex minus -.2ex}{1.5ex plus 0.3ex minus0.2ex}{\large\bf\raggedright}}
\def\subsection{\@startsection{subsection}{2}{\z@}{-1.8ex plus-0.5ex minus -.2ex}{0.8ex plus .2ex}{\normalsize\bf\raggedright}}
\def\subsubsection{\@startsection{subsubsection}{3}{\z@}{-1.5ex plus -0.5ex minus -.2ex}{0.5ex plus .2ex}{\normalsize\bf\raggedright}}
\def\paragraph{\@startsection{paragraph}{4}{\z@}{1.5ex plus 0.5ex minus .2ex}{-1em}{\normalsize\bf}}
\def\subparagraph{\@startsection{subparagraph}{5}{\z@}{1.5ex plus  0.5ex minus .2ex}{-1em}{\normalsize\bf}}

\makeatother

\skip\footins 9pt plus 4pt minus 2pt
\def\footnoterule{\kern-3pt \hrule width 12pc \kern 2.6pt }
\leftmargini\leftmargin \leftmarginii 2em
\makeatletter
\def\@listi{\leftmargin\leftmargini}
\def\@listii{\leftmargin\leftmarginii
   \labelwidth\leftmarginii\advance\labelwidth-\labelsep
   \topsep 2pt plus 1pt minus 0.5pt
   \parsep 1pt plus 0.5pt minus 0.5pt
   \itemsep \parsep}
\def\@listiii{\leftmargin\leftmarginiii
    \labelwidth\leftmarginiii\advance\labelwidth-\labelsep
    \topsep 1pt plus 0.5pt minus 0.5pt 
    \parsep \z@ \partopsep 0.5pt plus 0pt minus 0.5pt
    \itemsep \topsep}
\def\@listiv{\leftmargin\leftmarginiv
     \labelwidth\leftmarginiv\advance\labelwidth-\labelsep}
\def\@listv{\leftmargin\leftmarginv
     \labelwidth\leftmarginv\advance\labelwidth-\labelsep}
\def\@listvi{\leftmargin\leftmarginvi
     \labelwidth\leftmarginvi\advance\labelwidth-\labelsep}
\makeatother

\belowdisplayskip \abovedisplayskip
\makeatletter
\def\normalsize{\@setsize\normalsize{11pt}\xpt\@xpt}
\def\small{\@setsize\small{10pt}\ixpt\@ixpt}
\def\footnotesize{\@setsize\footnotesize{10pt}\ixpt\@ixpt}
\def\scriptsize{\@setsize\scriptsize{8pt}\viipt\@viipt}
\def\tiny{\@setsize\tiny{7pt}\vipt\@vipt}
\def\large{\@setsize\large{14pt}\xiipt\@xiipt}
\def\Large{\@setsize\Large{16pt}\xivpt\@xivpt}
\def\LARGE{\@setsize\LARGE{20pt}\xviipt\@xviipt}
\def\huge{\@setsize\huge{23pt}\xxpt\@xxpt}
\def\Huge{\@setsize\Huge{28pt}\xxvpt\@xxvpt}
\makeatother

\def\toptitlebar{\hrule height4pt\vskip .25in\vskip-\parskip}
\def\bottomtitlebar{\vskip .29in\vskip-\parskip\hrule height1pt\vskip .09in}

\begin{document}

\title{Retinal oscillations carry \\visual information to cortex}

\author{Kilian Koepsell\thanks{Redwood Center for Theoretical Neuroscience,
    University of California Berkeley, 132 Barker Hall, MC \#3190, Berkeley,
    CA 94720-3190, USA}, \hspace{.5em}
  Xin Wang\thanks{Neuroscience Graduate Program, University of Southern
    California, Los Angeles, CA, USA}, \hspace{.5em}
  Vishal Vaingankar\makebox[0pt]{$\,{}^\dagger$}, \hspace{.5em}
  Yichun Wei\makebox[0pt]{$\,{}^\dagger$}, \hspace{.5em}
  Qingbo Wang\makebox[0pt]{$\,{}^\dagger$}, \LINEBREAK
  Daniel L. Rathbun\thanks{Center for Neuroscience, University of California,
    Davis, CA, USA}, \hspace{.5em}
  W. Martin Usrey\makebox[0pt]{$\,{}^\ddagger$}, \hspace{.5em}
  Judith A. Hirsch\makebox[0pt]{$\,{}^\dagger$} \hspace{.5em} \textbf{\&} \hspace{.5em}
  Friedrich T. Sommer\makebox[0pt]{$\,{}^*$}}
\maketitle

\begin{abstract}
  Thalamic relay cells fire action potentials that transmit information from
  retina to cortex. The amount of information that spike trains encode is
  usually estimated from the precision of spike timing with respect to the
  stimulus. Sensory input, however, is only one factor that influences neural
  activity. For example, intrinsic dynamics, such as oscillations of networks
  of neurons, also modulate firing pattern. Here, we asked if retinal
  oscillations might help to convey information to neurons downstream.
  Specifically, we made whole-cell recordings from relay cells to reveal
  retinal inputs (EPSPs) and thalamic outputs (spikes) and analyzed these
  events with information theory. Our results show that thalamic spike trains
  operate as two multiplexed channels. One channel, which occupies a low
  frequency band ($<\!30$ Hz), is encoded by average firing rate with respect
  to the stimulus and carries information about local changes in the image
  over time. The other operates in the gamma frequency band (40-80 Hz) and is
  encoded by spike time relative to the retinal oscillations. Because these
  oscillations involve extensive areas of the retina, it is likely that the
  second channel transmits information about global features of the visual
  scene. At times, the second channel conveyed even more information than the
  first.
\end{abstract}

\section*{Introduction}

Thalamic relay cells transmit the information encoded in retinal firing rates
downstream to cortex. It is widely held that the amount of information that
retinal spikes carry is limited by the precision with which the firing rate
tracks changes in the stimulus, a framework of stimulus-locked rate
coding. Yet neural coding is not limited to stimulus-locked patterns of
response; the dynamics of intrinsic networks~\citep{Brivanlou1998,Meister1995}
also influences firing. Indeed, work in several modalities suggests that
information can be encoded by spike timing with respect to ongoing oscillatory
activity~\citep{Ahissar1990,Szwed2003,Friedrich2004,OKeefe1993,Montemurro2008}.

Oscillations in the firing rate of retinal ganglion cells are seen in species
as diverse as the frog and
cat~\citep{Arai2004,Heiss1965,Heiss1966,Laufer1967,Castelo-Branco1998}. A
natural question is whether these intrinsic retinal rhythms might provide
information to higher stages in the visual pathway. To address this subject,
we made whole-cell recordings {\em in vivo} from the cat's lateral geniculate
nucleus of the thalamus (LGN) during the presentation of natural movies. With
this technique, it was possible to detect both individual retinal inputs and
the spikes they evoke from single relay cells. In addition, we analyzed
extracellular recordings of retinal activity obtained in a separate laboratory
as a control. Thus far, our results show that oscillations in retinal inputs,
EPSPs, can be transmitted to cortex by thalamic outputs, spikes.

We next used information theory to explore how both external visual stimuli
and intrinsic rhythms modulate patterns of thalamic activity~\citep[for further
details, see][]{Koepsell2008a}. Our analyses showed that these two components
of the input a single thalamic neuron receives are multiplexed into two
parallel channels. One channel is encoded by average firing rate with respect
to the stimulus, stimulus-locked coding. It operates in the low frequency
band ($<\!30$ Hz) and carries information about sequential changes in the visual
stimulus. The second channel is encoded by the timing of individual spikes
relative to the retinal oscillations, oscillation-based coding. This channel
operates in the gamma-frequency band (40-80 Hz) and, because oscillations
involve distributed networks, is likely to convey information about
large-scale features or spatiotemporal context. Remarkably, the amount of
information in the second channel could match or even exceed that conveyed by
the first. Further we were able to reproduce this result with a simple model
of a relay cell. Thus, these two mulitplexed channels are not, in principle,
difficult to generate.

The presence of two information channels in the spike trains of single relay
cell offers substantial advantages for transmission of information to the
cortex. For example, dual channels could enhance robustness of the system to
noise since they offer two mechanisms for decoding spike trains. Moreover,
the second channel provides a conduit for novel information that is not
conveyed by stimulus-locked changes in firing rate alone.

\section*{Results}

The results were obtained from adult cats in two different laboratories, each
using different recording techniques, visual stimuli and anesthetics. The
main dataset includes whole-cell recordings from thalamic relay cells were
made from 15 subjects. A second dataset, that served as a control, included
extracellular recordings from retinal axons in the optic tract from 3
subjects.

\begin{figure}[ht]
\begin{center}
\includegraphics[width=.8\linewidth]{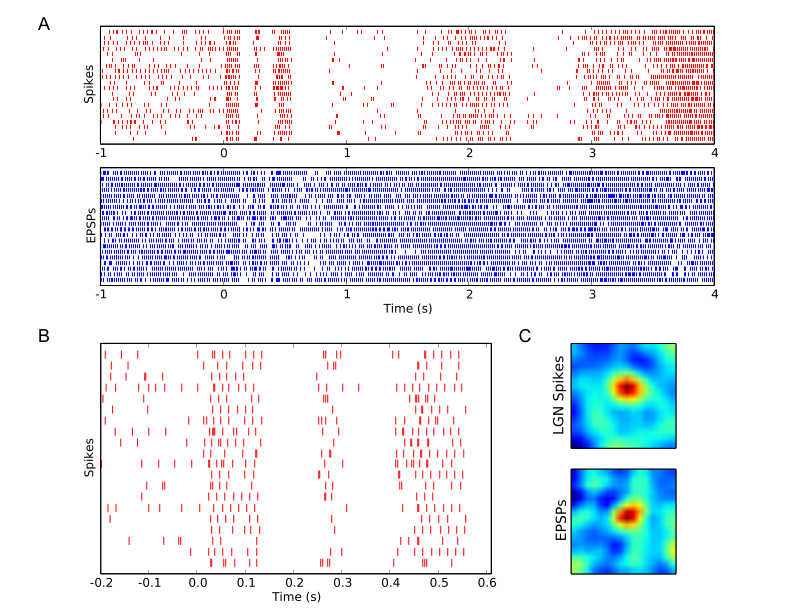} 
\end{center}
\caption{{\em Timing of retinogeniculate EPSPs and thalamic spikes recorded
    intracellularly from a single relay cell during the presentation of
    natural movies.}  (A)~Rasters of timings for spikes (red, top) and EPSPs
  (blue, bottom) in response to multiple trials of a movie clip that began at
  time $t\!=\!0$. (B)~Raster for spike timings around the movie onset with
  higher resolution than in A. (C)~Receptive fields for spikes (top) and EPSPs
  (bottom) mapped from responses to a movie; red indicates excitation to
  bright and blue excitation to dark.}
\label{fig:1}
\end{figure}

\subsection*{The temporal structure of inputs and outputs of single relay cells}

We focused our analyses on responses to natural movies, stimuli that reprise
features present in the environment. In order to examine the relationship
between the retinal inputs and thalamic outputs that these movies evoked, we
used cluster analysis. This method allowed us to label excitatory synaptic
potentials (EPSPs) in the intracellular signal and to separate these from
spikes. Results for a single neuron are illustrated as raster plots where time
points for spikes (Fig.~\ref{fig:1}A, top) are red and for EPSPs
(Fig.~\ref{fig:1}A, bottom) are blue. Each row shows the response to a 5~s
clip of the full stimulus. Raster plots show how inputs (EPSPs) and outputs
(spikes) tracked changes in the stimulus; their rates systematically sped and
slowed during repeated presentations of the same movie clip. An expanded view
of the spike trains shows that there was a fair amount of jitter between one
trial of the stimulus and the next; this variability is greater than that
recorded for full field flicker~\citep[data not shown, and
see][]{Liu2001,Reinagel2000,Eckhorn1975}. From the averaged rate over time we
were able to extract the receptive fields from the response~\citep{Wang2007}
for both the retinal inputs (Fig.~\ref{fig:1}C top) and the spikes
(Fig.~\ref{fig:1}C, bottom). Both of the receptive fields have the round
shape characteristic of ganglion cells. The shape of the receptive field of
the inputs~\citep{Usrey1998} is consistent with the interpretation that they
are retinogeniculate EPSPs fed forward from the retina rather than fed back
from the cortex. This conclusion is supported by the prominent size of the
events and the fast maintained firing rates~\citep{Frishman1983} (see
Fig.~\ref{fig:3}A). Not only are unitary cortical inputs rarely big enough to
visualize~\citep{Granseth2003} unless the membrane resistance is made larger
by blockade of potassium channels but cortical cells in layer 6 have very low
maintained rates~\citep{Gilbert1977}.

\begin{figure}[ht]
\begin{center}
\includegraphics[width=.8\linewidth]{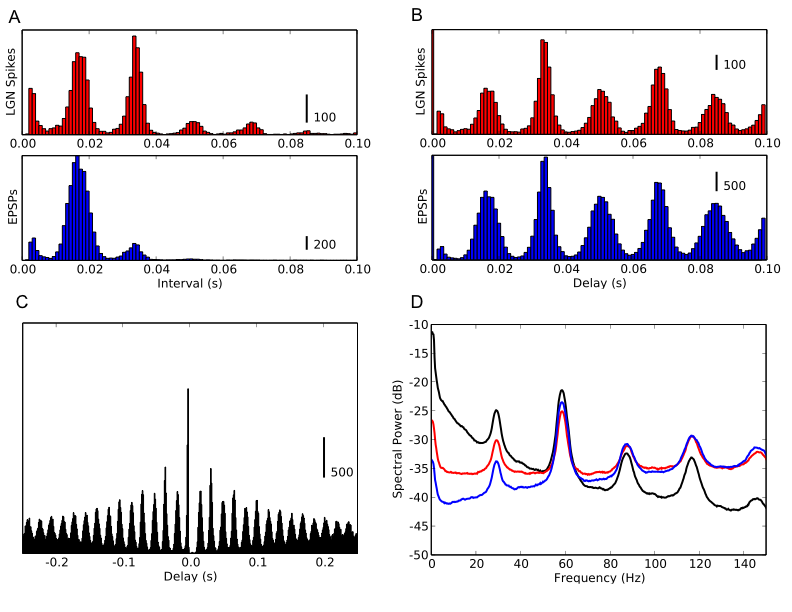} 
\end{center}
\caption{{\em Periodicity in timings of LGN spikes and EPSPs.}
  (A)~Time-interval histogram for spikes (red, top) and EPSPs (blue,
  bottom). (B)~Auto-correlation histogram for spikes (red, top) and EPSPs
  (blue, bottom). (C)~Cross-correlation histogram between EPSPs and spikes
  (spike at $t\!=\!0$). (D)~Power spectra of membrane potential (black),
  spikes (red) and EPSPs (blue) of same cell as in A-C, computed with
  multitaper method.}
\label{fig:2}
\end{figure}

\subsection*{Intrinsic rhythmic activity}

Neural activity is dictated by a combination of external stimuli and internal
dynamics. The event times shown in Fig.~\ref{fig:1} were plotted with respect
to the onset of the stimulus. We next plotted the data with respect to the
interval between the spikes or EPSPs as inter-spike intervals
(Fig.~\ref{fig:2}A) or autocorrelograms (2B). The multi-modal shapes of these
time interval histograms revealed additional temporal structure in the neural
responses; the tall peak at 17 ms and lesser peaks at multiples of that value
showed that both the EPSPs and spikes oscillated near 59 Hz,
Fig.~\ref{fig:2}A. Further, cross correlation of the two sets of events
revealed a sharp peak near zero (Fig.~\ref{fig:2}C), showing that thalamic
spikes followed individual retinal EPSPs with millisecond delays. The
oscillations are also visible in the power spectra for EPSPs (red), spikes
(blue), and the membrane potential (black), Fig.~\ref{fig:2}D. These
oscillations are not visible in the rasters shown in Fig.~\ref{fig:1},
because the intrinsic retinal rhythms are not synchronized with the
stimulus.

\begin{figure}[ht]
\begin{center}
\includegraphics[width=.8\linewidth]{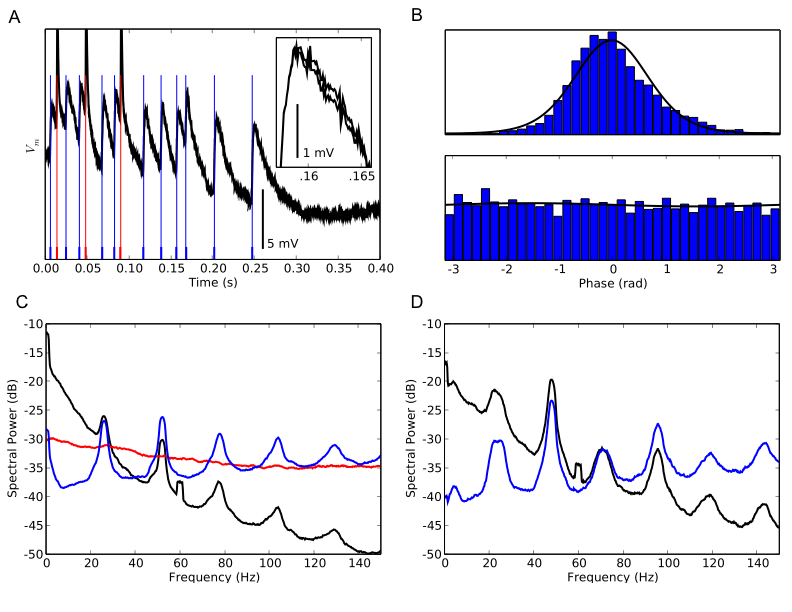} 
\end{center}
\caption{{\em Controls for stimulus and noise artifacts in recordings.}
  (A)~Example of membrane voltage from a noisy recording before (gray trace)
  and after (black trace) removal of line noise (60$\pm$0.1 Hz), see also
  magnified inset. The 60 Hz noise is visible, though far smaller than the
  biological signals. Vertical lines mark timings of detected spikes (red) and
  {EPSPs} (blue). (B)~Phase distribution of EPSP events relative to band
  passed (51.5$\pm$2 Hz) membrane potential (top) and relative to line
  noise (60$\pm$0.1 Hz) component (bottom). (C)~Spectra of membrane potential
  (black), spikes (red) and EPSPs (blue) during natural movie
  stimulation. (D)~Spectra obtained from recordings of the same cell as in A
  with the eyes of the cat closed.}
\label{fig:3}
\end{figure}

Since neural oscillations like that illustrated above often have frequencies
in the gamma range, which includes power line frequencies (50 or 60 Hz) there
is understandable concern that contamination of the biological signal can
introduce false rhythms. Such concerns stem from experience with extracellular
records in which spike heights often are on the order of tens of microvolts in
amplitude, orders of magnitude smaller than the intracellular signals studied
here.

Instances in which our recordings were contaminated with line noise provide
useful controls for separating neural signal from electrical interference. An
intracellular recording that included transients from the power line is
overlaid with blue and red vertical lines that indicate the timings of EPSPs
and spikes respectively, see Fig.~\ref{fig:3}A. The inset shows an overlay of
an EPSP marred by the artifactual transient and of the same EPSP after notch
filtering at 60 Hz. Except for the removal of the artifacts, the difference
between the shapes of the EPSPs is negligible.

A separate analysis compared the relative timing of neural events to the
oscillations we measured vs. their timing relative to the power line. If the
neural events, EPSPs and spikes, locked selectively to the biological rhythms,
then the distribution of event times plotted against phase of the intrinsic
oscillations should have a strong peak. By contrast, a similar plot of event
times with respect to the line frequency should be flat. These are exactly the
distributions we observed, see Fig.~\ref{fig:3}B.

Another way to examine the consequences of contamination by the line power is
to compare the power spectra of the raw signal to those of the event
trains. The contribution of line noise is visible as a narrow peak at 60 Hz
but is absent from the spectra of the EPSP and spike trains, see
Fig.~\ref{fig:3}C.

There is also a possibility that the oscillations might simply reflect
entrainment to the refresh rate of video display. We reduced this risk by
using rapid video refresh rates, above 140
Hz~\citep{Wollman1995,Williams2004,Butts2007}. This tactic was successful, as
shown by a comparison of recordings obtained during visual stimulation and
those made when the eyes were occluded and the monitor switched off, see
Fig.~\ref{fig:3}C and Fig.~\ref{fig:3}D. The spectra for the membrane
potential (black) and EPSPs (blue) were similar; the cell fired too few spikes
when the eye was closed to compute a spectrum.

\begin{figure}[ht]
\begin{center}
\includegraphics[width=.8\linewidth]{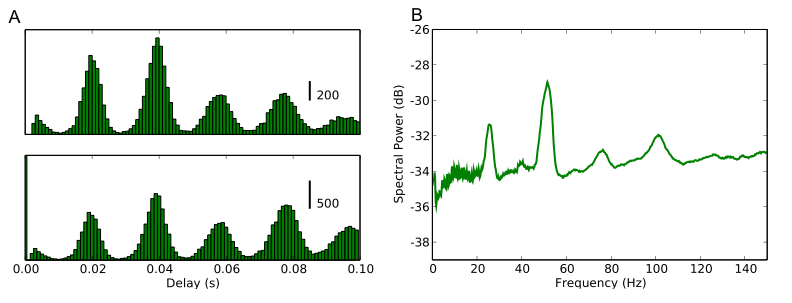} 
\end{center}
\caption{{\em Periodicity in spike timings of retinal ganglion cells.}
  (A)~Time-interval histogram (top) and auto-correlation histogram
  (bottom) for spikes of a retinal ganglion cell from extracellular recording
  in optic tract. (B)~Power spectrum of spike timings of the same cell
  as in A, computed with multitaper method.}
\label{fig:4}
\end{figure}

Finally, to rule out the possibility that the oscillations were unique to our
intracellular methods, we analyzed 20 extracellular recordings from
retinothalamic axons in the optic tract obtained in a second laboratory. Gamma
oscillations were present in recordings from this preparation, as depicted for
one cell, see Fig.~\ref{fig:4}.

\subsection*{Range of Oscillation Strengths}

How can even small contributions of oscillatory activity be quantified? So far
we have used simple measures that explore either the time domain (plots of
inter-spike intervals and autocorrelations) or the frequency domain (power
spectra). These measures provide a direct and intuitive means of displaying
prominent oscillations. But the number and height of the peaks in plots of the
inter-spike interval histograms or autocorrelograms lack the sensitivity to
reveal weak oscillations. Further, power spectra often contain spurious peaks
due to the refractoriness of the spiking process. Thus, we used the
oscillation score~\citep[$OS$, see Methods and][]{Muresan2008} to examine the
strength and frequency of oscillations in the gamma range for three datasets:
the whole cell recordings in current clamp we have discussed so far and two
additional control groups. One control group comprised recordings made in
voltage rather than current clamp mode (synaptic events are easier to detect
in voltage clamp). The second control group was made of the extracellular
recordings from the optic tract.

\begin{figure}[ht]
\begin{center}
\includegraphics[width=.8\linewidth]{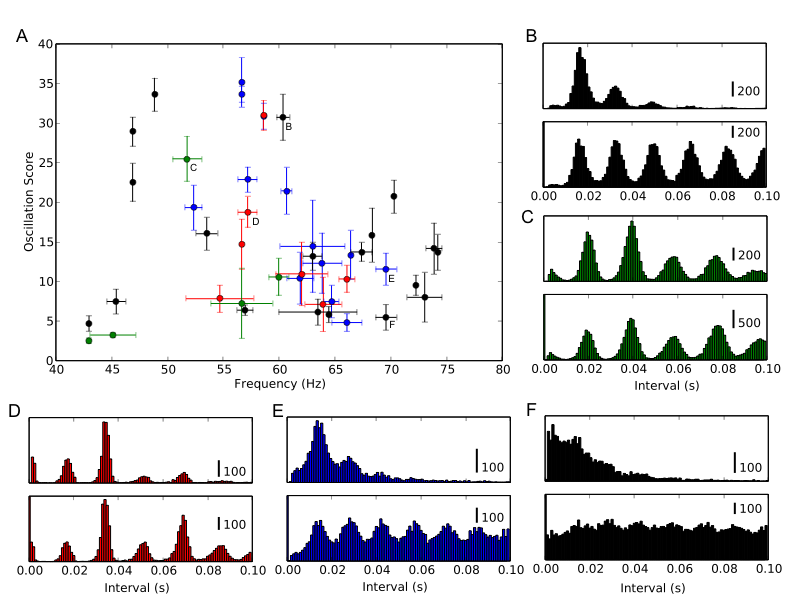} 
\end{center}
\caption{{\em Oscillations in population of all recorded cells.}
  (A)~Oscillation strength quantified by oscillation score as a function of
  oscillation frequency for all recorded cells that show oscillatory
  behavior. LGN spike outputs are shown in red, LGN EPSPs in blue, EPSCs from
  LGN voltage clamp recordings in black, and retinal ganglion spikes from
  optic tract recordings in green. The labeled points refer to the examples in
  panel B-F. (B)-(F)~Examples of spike interval histograms (top) and auto-
  correlation histograms (bottom) for cells with different oscillation
  score. Color convention as in A.}
\label{fig:5}
\end{figure}

A scatter plot of oscillation score against frequency showed a range of
oscillation strengths from weak to strong, see Fig.~\ref{fig:5}A. We included
all cells in the analysis that oscillated with a consistent frequency
($\sigma_f\!<\!4$~Hz across trials). Surrounding pairs of histograms of
inter-spike interval (top) and autocorrelation functions (bottom) correspond
to lettered points in the scatter plot and provide comparison of raw data from
which those scores were derived, Figs.~\ref{fig:5}B-F). Note that there is a
small but significant score for a cell whose inter-spike interval plot shows
little evidence of oscillatory activity (Fig.~\ref{fig:5}F).

The values plotted in Fig.~\ref{fig:5} are also summarized in
Table~\ref{tab:1}. We usually recorded in voltage clamp mode, but singled out
cells that seemed to oscillate for further analysis in current clamp. There is
modest discrepancy between the frequency of oscillations from the
extracellular retinal and intracellular thalamic records.  This difference
might reflect differences in sampling bias, visual stimuli (natural movies for
the intracellular experiments and m-sequences for the extracellular
experiments), or anesthetic (thiopental, a barbiturate, for the extracellular
experiments and, propofol, a new class of anesthetic, for the intracellular
experiments; note, however, that oscillations have been observed in awake
cats~\citep{Doty1964,Heiss1965,Heiss1966}). It is also possible that
differences reflect a circuit property of the thalamus; relay cells often
receive input from several ganglion cells~\citep{Hamos1985,Usrey1999}. If only
one of several inputs to a relay cell oscillated, we would have scored that
relay cell as oscillatory even if its other retinal inputs fired
aperiodically. There also was a larger percentage of oscillating EPSP than
spike trains; this difference reflects the limitation of the oscillation score
measure for very low spike rates.

\begin{table}[ht]
  \caption{Oscillations in different datasets}
  \begin{center}
  \begin{tabular}{| r | l || c | c | c |}
    \hline
    && \bf\parbox{8.5ex}{\# cells in\\data set} &
    \bf\parbox{15ex}{\begin{center}\# cells with\\consistent osc.\\($\boldsymbol{\sigma_f\!<\!4}$ Hz)\end{center}}&
    \bf\parbox{13ex}{\begin{center}\# cells with\\strong osc.\\($\boldsymbol{OS\!>\!15}$)\end{center}}\\
    \hline
    \hline
    \multirow{2}{7ex}{\begin{sideways}\parbox{10ex}{Current\\clamp\\recordings}\end{sideways}}
    & \bf\parbox{14ex}{$\phantom{x}$\\Retinothalamic EPSPs\\$\phantom{x}$} & 20 & 13 (65\%) & 7 (35\%) \\

    & \bf\parbox{10ex}{$\phantom{x}$\\Thalamic spikes\\$\phantom{x}$} & 20 & 6 (30\%) & 2 (10\%) \\
    \hline

    \multirow{2}{7ex}{\begin{sideways}\parbox{10ex}{Control\\recordings}\end{sideways}}
    & \bf\parbox{23ex}{$\phantom{x}$\\Retinothalamic EPSCs\\recorded only\\in voltage clamp\\$\phantom{x}$}
    & 30 & 19 (63\%) &  7 (23\%) \\
    & \bf\parbox{23ex}{$\phantom{x}$\\Retinal spikes\\from the optic tract\\$\phantom{x}$} & 20 & 5 (25\%) & 1 (5\%) \\
    \hline
  \end{tabular}
  \end{center}
  \label{tab:1}
\end{table}

\begin{figure}[ht]
\begin{center}
\includegraphics[width=.8\linewidth]{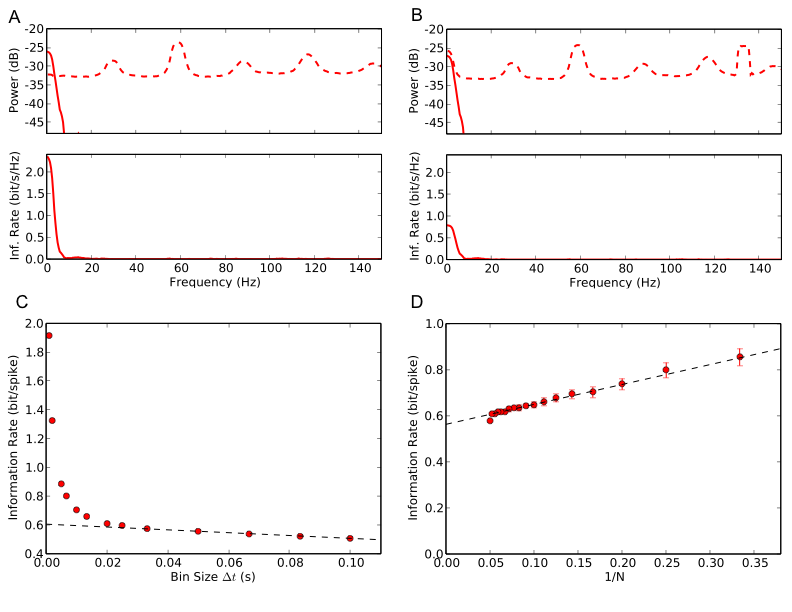} 
\end{center}
\caption{{\em Estimates for information rate of LGN spikes.}  (A)~Power
  spectrum (top) of LGN spike response decomposed into signal (red line) and
  noise (dashed line) and spectral information rate (bottom) as upper bound
  for the transmitted information; the total information rate given by the
  area under the curves is 8.3~bit/s. With a mean spike rate of 19~spikes/s,
  this corresponds to 0.4~bit/spike. (B)~Power spectrum (top) of LGN spike
  response decomposed into signal (red line) and noise (dashed line) with the
  stimulus reconstruction method as lower bound for the transmitted
  information; the estimated information rate is 5.7~bit/s or 0.3~bit/spike,
  similar to values measured
  previously~\citep{Liu2001,Reinagel2000,Eckhorn1975}. (C)~Information per
  spike as a function of bin width $\Delta t$. Linear extrapolation ($\Delta t
  \rightarrow 0$) yields an information rate of 0.58~bit/spike. (D)~Direct
  information estimate as a function of number of trials N. Linear
  extrapolation ($\Delta t,\,1/N\rightarrow 0$) yields 0.56~bit/spike.}
\label{fig:6}
\end{figure}

\subsection*{Analyses of information content}

Having established the presence of an oscillatory component of
retinogeniculate responses, we began our analyses of the information content
in retinal and thalamic event trains. The first analyses were made from the
perspective of stimulus-locked coding, using the current clamp recordings. We
estimated lower and upper bounds for the amount of information that could be
transmitted by changes in event rate that recurred across repeated
presentations of the same stimulus. In order to obtain each bound, it was
necessary to use separate methods~\citep{Bialek1991}. To establish the upper
bound, we decomposed the power spectrum of the spike trains into two
components. One component represented the part of the response that was
consistent across stimulus trials, typically equated with the signal
(Fig.~\ref{fig:6}A, top, solid curve) and the other corresponded to the
variation in response across trials, usually taken as noise,
(Fig.~\ref{fig:6}A, top, dotted curve). The information rate, calculated from
the area under the solid curve in Fig.~\ref{fig:6}A bottom, was 0.4
bit/spike. We established the lower bound by determining how well the visual
stimulus could be reconstructed from the convolution of the cellular response
with the receptive field (Fig.~\ref{fig:6}B, and Methods). For this case, the
signal is the reconstruction of the receptive field and the noise is the
deviation of the reconstruction and the stimulus. The information rate
calculated from this second method was 0.3 bit/spike, Fig.~\ref{fig:6}B,
bottom. The values for upper and lower bounds are similar to those reported
previously~\citep{Liu2001,Reinagel2000,Eckhorn1975}.

The information rates estimated above were made with the widely used entropy
methods for a Gaussian information channel, which assumes that signal and
noise are Gaussian~\citep{Borst1999}. We also estimated information rates using
the direct method for single spike information~\citep{Brenner2000}, which is
valid for arbitrary distributions but assumes that single spikes encode
information independently (Figs.~\ref{fig:6}C, D). The direct method, however,
requires so much data to make estimates at fine times scales that it is common
to obtain estimates by extrapolating from values obtained at coarser
time scales (wider bins, Fig.~\ref{fig:6}C). The estimates obtained using the
entropy method, 0.4 bit/s and the direct method, 0.58 bit/s were very
close. The similarity of these two values, each obtained with a different,
complementary, method, suggests that they reflect the true information rate.

The results plotted in Fig.~\ref{fig:6} illustrate an important aspect of how
relay cells use stimulus-locked rate coding to encode visual signals. The
information about temporal changes in the stimulus peaked at low frequency and
dropped sharply at a cut-off frequency below 30 Hz (Fig.~\ref{fig:6}A
and~\ref{fig:6}B). The shape of this distribution reflects not only the
intrinsic properties of thalamic circuits but also the statistics of natural
images, which are skewed to low spatial and temporal
frequencies~\citep{Ruderman1994}. This finding is consistent with previous
studies that used natural scenes~\citep{Dan1996} rather than flickering,
full-field stimuli~\citep{Reinagel2000,Liu2001}.

\begin{figure}[ht]
\begin{center}
\includegraphics[width=.8\linewidth]{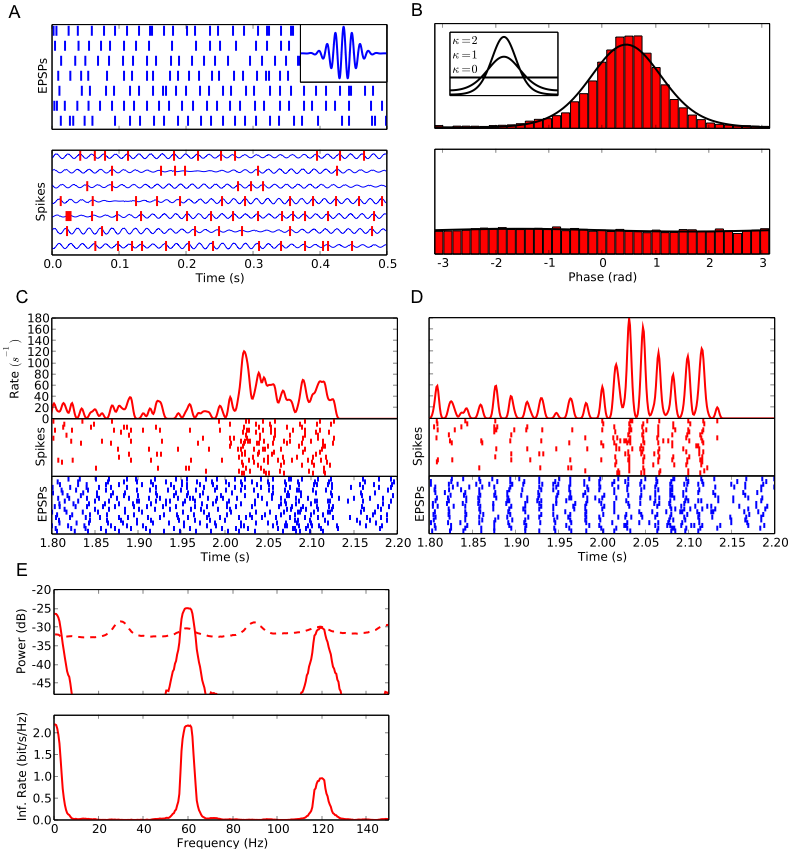} 
\end{center}
\caption{{\em Adjusting response latency based on the phase of ongoing
    oscillations reduces temporal jitter in spike timings across trials.}
  (A)~EPSP trains (top) and spikes (bottom, red) for 7 trials, analytical
  signal (bottom, blue curve) computed from EPSP trains of each trial by
  filtering with a Morlet wavelet (inset). (B)~Histogram of spike phases (top)
  and shift predictor (bottom) for an oscillation frequency of
  59~Hz. (C)~Response recorded with fixed latency to stimulus onset. Averaged
  spike rate (red curve) and rasters for spikes (red) and EPSPs (blue) for 20
  trials of a movie clip. Spikes rasters were smoothed with Gaussian window
  ($\sigma\!=\!2$~ms) before averaging. (D)~Responses corrected for latency
  variations up to $\pm10$~ms by using periodicity in the ongoing activity
  that preceded stimulus onset; conventions as in A. (E)~Power spectrum (top)
  of de-jittered spike train decomposed into signal (solid line) and noise
  (dashed line) for single cell and spectral information rate
  (bottom). De-jittering increased the total information from 0.3 -
  0.4~bit/spike (Fig.~\ref{fig:6}A/B) to 1.6~bit/spike.}
\label{fig:7}
\end{figure}

The prominence of the oscillations in the "noise" motivated us to ask if it
might actually encode information hidden from conventional analyses.
Answering this question required that we measure the degree to which the
oscillations contributed to the variance of the stimulus-locked response. We
estimated the instantaneous phase of the retinal oscillations (EPSPs) as the
complex angle of an analytical signal calculated by convolving a Morlet
wavelet (Fig.~\ref{fig:7}A top, inset) with the event train (blue ticks in
Fig.~\ref{fig:7}A); and see Methods. The resulting waveform (blue ripple,
Fig.~\ref{fig:7}A bottom) was then used to determine the phases at which the
relay cell spiked (red ticks, Fig.~\ref{fig:7}A, bottom). This analysis showed
that the phase locking between retina and thalamus was strong, as made clear
by the tall peak in the phase histogram plotted in Fig.~\ref{fig:7}B
top. There was no evidence, however, that each new trial of the stimulus set
the absolute phase of the neural response: when we used the estimated phase
for inputs in one trial to determine the phase of spikes in the next, the
resulting distribution (the shift predictor) was flat (Fig.~\ref{fig:7}B,
bottom).

We further quantified the degree to which the thalamic spikes locked to the
retinal inputs with a concentration parameter $\kappa$ of a von Mises
distribution that we fitted to each phase histogram (see Fig.~\ref{fig:7}B,
Methods). The concentration parameter is zero for a flat phase distribution
and increases with increasing degree of phase locking between retinal input
and thalamic output. Thus fits with the tallest peaks indicate the highest
degree of phase locking and the most reliable transmission of oscillations in
the inputs (Fig.~\ref{fig:7}B, inset). The value for this cell was 2.3 and the
range for all cells was 0 -- 3.1.

Next we used our method of estimating the phases of the retinal oscillations
to align the phases of EPSPs over different trials. This allowed us to explore
how the random phases of oscillations across trials might have introduced
jitter and hence decreased precision in the stimulus-locked response. To align
the phase of the responses across trials, we measured the local phase of
ongoing activity recorded just before each repeat and adjusted that phase to
match the mean phase (see Methods). The alignment to the instantaneous phase
reduced the jitter in the cross-trial latencies to a striking
extent. Remarkably, what had seemed to be randomly distributed events in the
actual recordings (Fig.~\ref{fig:7}C) assumed temporally precise patterns in
the de-jittered traces (Fig.~\ref{fig:7}D). Much of the variation across
trials that had seemed like random jitter had come from periodic activity. To
address directly the question of whether the oscillatory activity in the gamma
band could be used to transmit visual information, we aligned the phases of
the EPSPs within the complete dataset. Rather than realigning phases at the
start of each trial, as above, we made an alignment each time the local phase
of the response deviated markedly from the reference (see Methods); the power
spectrum of the de-jittered records, decomposed into signal and noise, is
plotted in Fig.~\ref{fig:7}E, top. Afterwards, we analyzed the de-jittered
dataset by using information theory just as we had done for the raw
recordings. The de-jittering exposed additional bands near 60 Hz and 120 Hz
that had high signal to noise ratios, Fig.~\ref{fig:7}E bottom, but left the
low frequency, 30Hz, band used for stimulus- locked coding in tact. When the
additional bands were taken into account, the upper bound on the rate of
information available from the spike train quadrupled; it grew from 0.4 to 1.6
bit/spike. This gain in information is possible because the band carrying
stimulus locked information is separate from the bands that carry the
oscillation based information. If spectra of the extrinsic and intrinsically
patterns of activity had overlapped, then oscillations would have interfered
with the information transmitted by the thalamic spike train.
 
\begin{figure}[ht]
\begin{center}
\includegraphics[width=.8\linewidth]{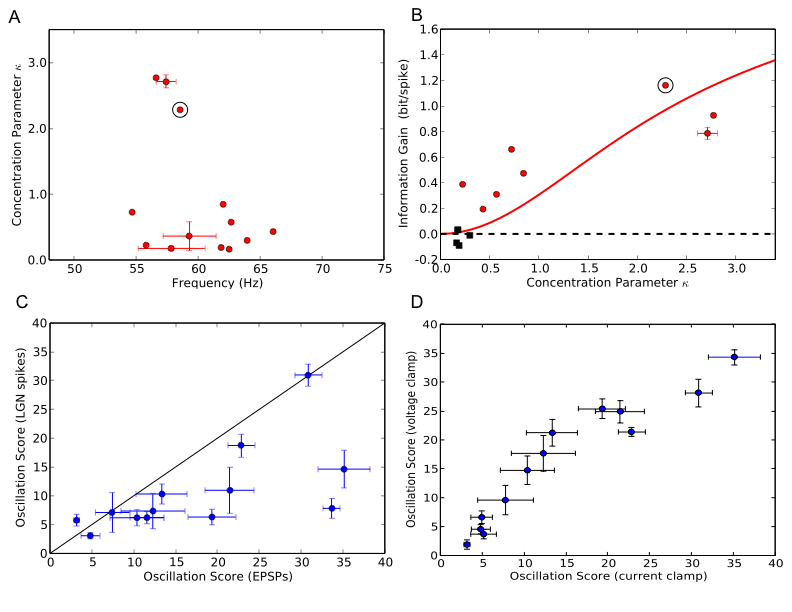}
\end{center}
\caption{{\em Population results for phase locking, information gain and
    oscillation score.} (A)~Phase-locking of thalamic spikes for all (13/20)
  cells that had oscillating EPSP trains (see Table~\ref{tab:1}). Error bars
  indicate standard deviation when multiple movies were presented; circled
  point denotes the cell analyzed in
  Figs.~\ref{fig:1},\ref{fig:2},\ref{fig:6},\ref{fig:7}. (B)~Gain in
  information rate after de- jittering plotted against concentration
  parameter~$\kappa$; conventions as in A. For 8 of 20 cells (red points), the
  information increased significantly ($p\!<\!0.05$, computed using
  permutation test) after de-jittering of the spike train (see A). Red curve
  depicts gain in transmitted information predicted by our model
  (Fig.~\ref{fig:9}) for variable~$\kappa$. (C)~Oscillation score of LGN
  outputs (spikes) as a function of oscillation score of LGN inputs
  (\mbox{EPSPs}). (D)~Oscillation scores of inputs to LGN cells (EPSCs, EPSPs):
  score of the voltage clamp recording shown as function of the score of the
  current clamp recording.}
\label{fig:8}
\end{figure}

Across the population, the amount of information available in the second
channel depended on the strength of the retinal oscillations. There was
pronounced phase-locking of retinal inputs to a wide range of gamma
oscillations, as shown in a plot of concentration parameter against frequency
(Fig.~\ref{fig:8}A). The gain in information after de-jittering was large for
cells with strong oscillations and commensurately smaller for cells with
weaker oscillations, red points in Fig.~\ref{fig:8}B. Here, as for
Figs.~\ref{fig:6}A and~\ref{fig:7}E, we estimated the information using the
entropy method for the Gaussian information channel. We chose this method
because temporal realignments of the spikes were so small, $<20$~ms, that the
amount of data required for the more general direct method (see Methods) was
unfeasibly large (see Fig.~\ref{fig:6}C). The red curve is a prediction of
information gain as a function of the concentration parameter; it was
generated with a computational model we describe below and calculated with the
direct method. Panel 8C explores the relationship between oscillation scores
for EPSPS vs. the spikes they evoke. Most points fall on or below the line of
unity slope; as mentioned earlier, the relatively reduced scores for the
spikes reflect limitations in the oscillation score measure for low spikes
rates in our recordings. Last, we obtained voltage-clamp recordings from many
cells; the oscillation scores calculated for the EPSCs in these data are
similar to those obtained from recordings made in current clamp
(Fig.~\ref{fig:8}D).
 
\begin{figure}[ht]
\begin{center}
\includegraphics[width=.8\linewidth]{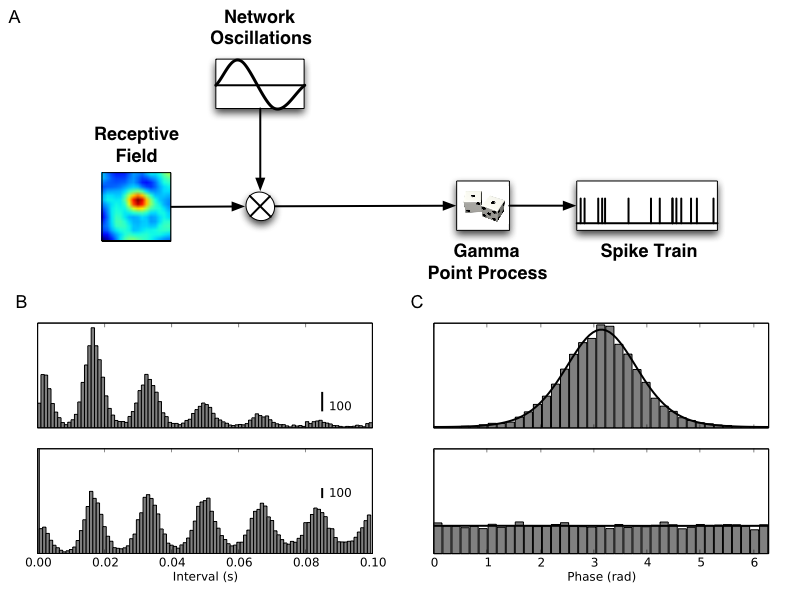} 
\end{center}
\caption{{\em Quasi-periodic gamma (QPG) model reproduces spike timing
    statistics and phase distribution.} (A)~QPG model of the thalamic neuron
  that predicts spike statistics and information rate: Spike times are
  described by inhomogeneous gamma process. Spike rate is the product of two
  signals, the visual response evoked through the receptive field and a
  periodic signal that simulates ongoing retinal activity. (B)~Simulated spike
  interval distribution (top) and auto-correlation histogram
  (bottom). (C)~Simulated spike phase distribution (top) and shift predictor
  (bottom).}
\label{fig:9}
\end{figure}

\subsection*{Modeling the multiplexed channels}

\begin{figure}[ht]
\begin{center}
\includegraphics[width=.8\linewidth]{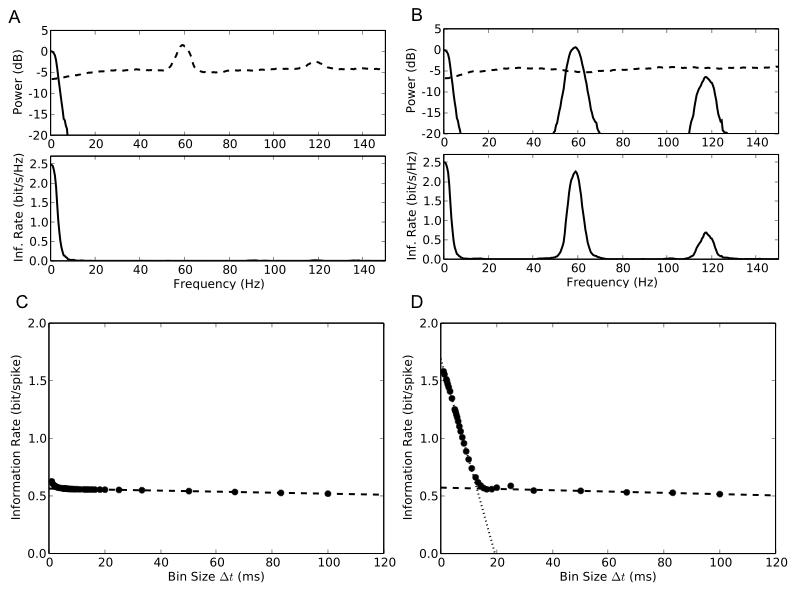} 
\end{center}
\caption{{\em Information rate for QPG model.}  (A)~Simulation with random
  oscillation phase across different trials: Signal spectrum (top, solid
  line), noise spectrum (top, dashed line) and information estimate
  (bottom). (B)~Simulation with oscillation phase aligned across different
  trials: Signal spectrum (top, solid line), noise spectrum (top, dashed line)
  and information estimate (bottom). (C)~Information estimate with oscillation
  phase randomized across trials (0.6 bit/spike). (D)~Information estimate
  with oscillation phase aligned across trials increases from 0.6 bit/spike to
  1.8~bit/spike if small bin width ($<20$~ms) are taken into account.}
\label{fig:10}
\end{figure}

Thus far, our analyses suggest that many thalamic relay cells convey
information about natural stimuli via spike trains that are multiplexed
between two channels that operate on different time scales. One channel
encodes information by modulations in spike rate determined by changes in the
stimulus and uses the frequency band below 30 Hz. A second channel utilizes
spike timing to convey information about ongoing retinal activity. This
channel uses higher, gamma band, frequencies. In order to understand how this
dual mode of transmission might be formed, we built a simple
model~\citep[Fig.~\ref{fig:9}A, see Methods and][]{Koepsell2008a}. Thalamic
spikes were generated by an inhomogeneous gamma process whose rate was
determined by a combination of two signals, one that corresponded to visual
input filtered by the receptive field and another that represented retinal
oscillations. The signals were combined by multiplication to reproduce the
amplitude modulation seen in event rates of the de-jittered recording
(i.e. Fig.~\ref{fig:7}D). Although our model had only four free parameters,
the order of the gamma process, the concentration parameter, the oscillation
frequency and the bandwidth, it was able to reproduce the key aspects of the
results. These features include inter-spike intervals and phase-locking
(compare Fig.~\ref{fig:9}B with 2B) and the power spectra and information
content of the raw (compare Fig.~\ref{fig:6} and~\ref{fig:10}A, and C)
de-jittered spike trains (compare Figs.~\ref{fig:7}E with~\ref{fig:10}B). The
simplicity of our model suggests that complicated mechanisms are not required
to generate the multiplexed channels.

\section*{Discussion}

We provide experimental evidence that individual retinal ganglion cells
multiplex two streams of information via the thalamus to cortex. One stream is
well known and encodes visual information by changes in firing rate that are
time-locked to external visual stimuli. The second, novel, channel encodes
information by aligning spike timing to intrinsic retinal oscillations. We
were able to expose this additional channel by developing a technique for
measuring the phase alignment of spikes fired by single relay cells to the
local phase of oscillations in presynaptic retinal input
input~\citep[see][]{Koepsell2008a}. Because this oscillation-based channel
operates in a frequency band separate from that containing the stimulus-locked
channel, it adds to the total amount of information transmitted to cortex. The
amount of extra information the second channel provides increases as a
function of oscillation strength and can as much as triple the number of bits
that each spike carries. Further these results are easily reproduced by a
simple computational model.

What information might the second channel encode? It is likely, that the
oscillation based channel transmits contextual information about the stimulus.
Oscillatory activity in retina is generated by the coordinated activity of
distributed networks that span large regions of retinal, hence visual,
space. Previous studies have shown that extended visual contours can
synchronize oscillations among distant ganglion
cells~\citep{Neuenschwander1996,Stephens2006}. Thus, global features might
modulate the frequency or amplitude of retinal oscillations or the alignment
of thalamic spikes to the phase of the retinal oscillations.  Hence, the
oscillation-based channel might serve to convey large-scale information such
as the overall gist of a scene~\citep{Navon1977,Torralba2003}. The information
that this channel transmits might be unique or might complement information
about spatial structure that is available in the relative timing of action
potentials fired by a single neuron~\citep{McClurkin1991,Klein1992,Reich1997}
or differences in the relative timing of parallel spike trains with respect to
visually evoked response latency~\citep{Gollisch2008}. The prediction that
oscillations convey such contextual information is supported by recent
experiments in frog. Blockade of retinal oscillations abolishes escape
behavior elicited by large stimuli that mimic predators but does not impair
detection of punctuate objects that resemble prey~\citep{Ishikane2005}. Thus
the use of oscillations to encode large-scale aspects of the visual scene
might be evolutionarily conserved in the visual system.

If, however, the oscillations were not modulated by the visual stimulus, as in
our model, they would still benefit visual processing. For example,
synchronization due to oscillations can reduce the probability of errors made
in encoding local features~\citep{Kenyon2004}. As well, they might help
transmit information effectively to the cortex, as follows. Retinal ganglion
cells often diverge to synapse with multiple relay
cells~\citep{Hamos1985,Usrey1999}. In turn, relay cells that receive common
retinal input often synapse with the same cortical target. The impact of each
spike a relay cell fires depends on physiological context: coincident inputs
are more likely to trigger a cortical spike than those that arrive at
intervals longer than a few ms~\citep{Usrey1998,Bruno2006}. Thus, oscillations
might synchronize the arrival of converging thalamic inputs onto a common
cortical target.

The presence of two channels that carry information could provide further
advantage by conveying redundant information about the visual stimulus in
separate frequency bands, as follows.  The first copy is encoded by a
straightforward mechanism, stimulus-locked changes in spike rate. To explain
how the second copy is conveyed, we use the analogy of AM radio
transmission. Here the visual signal modulates the amplitude of the higher
frequency carrier, which, in this case, is the gamma oscillation. Thus, the
visual stimulus is encoded redundantly, in two separate frequency bands of the
thalamic spike train.

This redundancy increases robustness to noise since it provides two
alternatives for cortex to decode the spike train, low passing or band
passing. The low passed information would be read by conventional mechanisms
of synaptic integration. The band-pass receiver is formed by cortical
oscillations in the gamma
band~\citep{Nowak1997,Hutcheon2000,Fellous2001}. Thalamic volleys that arrive
nearer the peaks of the cortical oscillations, when cortical neurons are most
depolarized, would be most likely to evoke spikes. This same scenario has been
suggested to explain how oscillations propagate active from one cortical area
to the next. Since the strength and frequency of gamma oscillations in cortex
is increased by attention, the contributions of the novel channel might be
enhanced during times of heightened vigilance to visual
signals~\citep{Fries2007}.

\section*{Methods}

\subsection*{Preparation, stimulation and recording for whole-cell experiments}

Adult cats (1.5 - 3.5 kg) were prepared as described
earlier~\citep{Hirsch1998} except anesthesia was maintained with propofol and
sufentanil. Whole-cell recordings of the membrane voltage or current were made
with dye filled pipettes from 17 cells in 9 animals using standard
techniques~\citep{Hirsch1998} (Axopatch 200A amplifier, Axon Instruments,
Inc., Union City, CA) and digitized at 10 kHz (power1401 data acquisition
system, Cambridge Electronic Design Ltd., Cambridge, UK). The stimuli were
various natural movies (30 s duration) that were repeated 5-50 times. The
movies were displayed at 19-50 frames per second on a video monitor (refresh
rate 133-160 Hz) by means of a stimulus generator (VSG2/5 or ViSaGe, Cambridge
Research Systems Ltd., Rochester, UK). Housing, surgical and recording
procedures were in accordance with the National Institutes of Health
guidelines and the University of Southern California Institutional Animal Care
and Use Committee.

\subsection*{Preparation, stimulation and recording for extracellular
 experiments}

Adult cats were prepared and anesthetized as described
earlier~\citep{Usrey1998}. The stimuli were white noise \citep[m-sequencs,
see][]{Usrey1998} created with a VSG2/5 visual stimulus generator (Cambridge
Research Systems Ltd., Rochester, UK) and updated at the 140 Hz refresh rate
of the video display. Spike trains were digitized at 20 kHz (power1401 data
acquisition system, Cambridge Electronic Design Ltd., Cambridge, UK) and
stored for further analysis. All procedures conformed to NIH guidelines and
were approved by the institutional Animal Care and Use Committee at the
University of California, Davis.

\subsection*{Event detection and sorting}

Potential events (spikes, EPSPs and false positives) were detected as
zero-crossings (from positive to negative) in the second derivative of the
intracellular signal. Spikes were distinguished from EPSPs by a threshold
criterion. A clustering algorithm~\citep{Harris2000} was used to separate
EPSPs from false positives; each event was characterized using a short segment
(1.5ms) of the second derivative. The first three principal components (in the
space of the event-centered second derivative) were used as features in the
clustering procedure; then clusters that corresponded clearly to retinal EPSPs
as determined by visual inspection were selected for further
analysis~\citep{Wang2007}. The extracted spike trains of the LGN cell (red)
and the event trains of the synaptic inputs (EPSPs) corresponding to retinal
spikes (blue) are displayed in Fig.~\ref{fig:1}A,B.

\subsection*{Estimating receptive fields}

We estimated the spatio-temporal receptive fields of the relay cells by using a
model that predicted neural responses to the natural movies
(Fig.~\ref{fig:1}C). Specifically we used regularized gradient
descent~\citep{Machens2004} to optimize a linear convolution kernel with
respect to the quadratic error between predicted and empirical response. The
firing rates used to calculate the receptive fields were estimated with a
temporal Gaussian filter, 25 ms half-width and spontaneous rates were taken as
the average rate across the entire recording~\citep{Wang2007}.

\subsection*{Estimating power spectra and information rate}
\label{sec:inf}

Upper and lower bounds on the information rate in Fig.~\ref{fig:6},
Figs.~\ref{fig:7}E and Fig.~\ref{fig:10} were estimated by the dynamic
Gaussian channel based on the power spectrum and double checked with the
direct method. We used a multitaper method~\citep{Jarvis2001}, with 5 tapers,
on non-overlapping windows to estimate spectral power
(Fig.~\ref{fig:2}D). These methods were based on different assumptions about
the statistics of the stimulus and the response as well as the neural model
\citep[for overview, see][]{Borst1999}.

\subsubsection*{Estimating information rate: the Gaussian channel}

When the signal has Gaussian statistics and the noise is additive and
Gaussian, information rates can be calculated as~\citep{Bialek1991,Rieke1999}:
\begin{equation}\label{eq:inf-gauss}
I_G = \frac{1}{2} \int_{-\infty}^{\infty}\frac{d\omega}{2\pi} \log\left[
  1+ SNR(\omega) \right]\, \mbox{bit/s},
\end{equation}
where $SNR(\omega) = S(\omega)/N(\omega)$ is the signal-to-noise ratio,
computed in the frequency domain from the spectrum of the signal $S(\omega)$
and the spectrum of the noise $N(\omega)$. When different definitions of
signal and noise were used, this formula yielded estimates for the upper and
lower bounds for the information rate, as below:

\vskip 5pt\noindent
\textit{Upper bound}

\noindent
The signal was defined as the component of the neural response that contained
information about the stimulus (determined by cross-trial average of spike
trains). The noise was defined as the deviation of individual trials from the
average. Power spectra were estimated using the multi-taper method. Errors due
to finite sample size ($N$) were corrected by assuming that the power of
Gaussian noise decreases as $1/N$ in the average across
trials~\citep{Sahani2003}. With these definitions, the formula for the
Gaussian channel gave estimates of an upper bound for the information that
could be transmitted by stimulus-locked coding (see also Fig.~\ref{fig:6}A,
Fig.~\ref{fig:7}E, Figs.~\ref{fig:10}A,B).

\vskip 5pt\noindent
\textit{Lower bound (Reconstruction method)}

\noindent
The signal was defined as the stimulus (the movie) and the noise as the
deviation between the actual stimulus and one that was reconstructed as
follows. To reconstruct the stimulus, the neural spike train was convolved
with a receptive field estimated from responses to a different movie (see
above). With these definitions, the formula for the Gaussian channel gave a
lower bound for the information that stimulus-locked coding could transmit
(see Fig.~\ref{fig:6}B). Note that the estimate of the lower bound is tight
only if the linear model used to reconstruct the stimulus captures the
response of the neuron well.

\subsubsection*{Using the direct method to estimate information rates}

So far, we have made that assumption that both the signal and noise are
Gaussian. If, however, the noise is not additive nor Gaussian, then $I_G$ does
not guarantee an upper bound. Similarly, if the signal is not Gaussian, then
$I_G$ does not guarantee a lower bound. Therefore, to assess how well the
Gaussian channel was able to capture the rates of information that relay cells
actually transmitted, we used a technique called the ``{\em direct method}''.
This method does not rely on Gaussian statistics or any given neural model to
estimate the total entropy in a neuron's response. If it is assumed that all
information is conveyed by single spikes rather than firing patterns, then the
information that each spike transmits is given by the
formula~\citep{Brenner2000}
\begin{equation}\label{eq:inf-direct}
I = \frac{1}{T} \int_{0}^{T}dt \frac{r(t)}{\overline{r}} \log_2\left[
  \frac{r(t)}{\overline{r}} \right]\, \mbox{bit/spike},
\end{equation}
where $r(t)$ is the spike rate and $\overline{r}$ is the mean rate averaged over
the whole recording time $T$. 

The direct method has a weakness, however; the accuracy of the estimate it
provides depends on the bin width, $\Delta t$, used to compute the integral in
the equation directly above.  Specifically, the estimate converges to the true
entropy only asymptotically (limit of zero bin width and infinite number of
trials).  Thus, narrow bins and finite data result in a pronounced
overestimation of the amount of information that a cell transmits, see
Fig.~\ref{fig:6}C, red dots.

The estimate can be improved by a linear extrapolation ($\Delta
t\rightarrow0$) of the values for larger bin sizes, Fig.~\ref{fig:6}C, dashed
line. A similar extrapolation must be made for estimating information with an
infinite number of trials, see Fig.~\ref{fig:6}D. The resulting value of 0.58
bit/spike was similar to the estimates obtained using the Gaussian channel
(upper bound: 0.44 bit/spike, lower bound 0.30 bit/spike). Two factors
probably account for the overestimation produced by the direct method. First, the
number of trials (20) was limited; increasing the number of trials ($N$) by
linear extrapolation ($\Delta t,\,1/N\rightarrow 0$) yielded 0.55 bit/spike
(Fig.~\ref{fig:6}D, dashed line). Second, the direct estimate did not remove
redundancies due to correlations between single spikes.

\subsection*{Oscillation score}

Muresan et. al~\citep{Muresan2008} used both the time and frequency domain to
devise a new metric for the strength and frequency of oscillations. Their
method to compute an `oscillation score' ($OS$) applies a Fourier transform on
an autocorrelation that is smoothed and whose central peak, including adjacent
troughs due to the refractory period, is subtracted; these precautions remove
confounds from the refractory period.

\subsection*{Quantification of the phase locking of spikes}

To characterize periodicity in trains of unitary events (EPSPs, spikes) it was
necessary to estimate frequency and phase at each point in time. Hence, we
computed the complex analytic signal $A(t)\!=\!A_0(t)\exp(i\phi(t))$ by
convolving the EPSP event train with a complex Morlet wavelet 
$$
w (t,f) = C \exp(2\pi i f t)\exp(-t^2/2\sigma_t^2)
$$
centered at a frequency $f$ with temporal width $\sigma_t$ and normalization
factor $C$ (Fig.~\ref{fig:7}A). The amplitude $A_0$ of the analytic signal
corresponded to the local power in the frequency band centered at $f$ with
bandwidth $\sigma_f\!=\!1/(2\pi\sigma_t)$. The bandwidth we chose,
$\sigma_f\!=\!2$~Hz, corresponded to the width of the peak in the power
spectrum and a temporal width of $\sigma_t\!=\!80$~ms. The instantaneous phase
$\phi(t)$ was estimated as the complex angle of the analytic signal and was
used to measure the distribution of phases for membrane events
(Fig.~\ref{fig:3}B, Fig.~\ref{fig:7}B, Fig.~\ref{fig:9}C). The phase
distribution of spikes (Fig.~\ref{fig:7}B) was fitted with a von Mises (or
cyclic Gaussian) distribution
$$
M(\phi| \kappa, \mu) = e^{\kappa cos(\phi - \mu)}/(2 \pi I_0(\kappa))\,.
$$
The mean phase $\mu$ was computed from the first trigonometric moment of the
spike distribution
$$
\langle \exp(i\phi) \rangle = \frac{1}{N}\sum_{n=1}^N\exp(i\phi(t_n))
= r \exp(i\mu)\,.
$$
The concentration parameter $\kappa$ characterizing the width of the von Mises
distributiion was obtained by numerical solution of the equation
$I_1(\kappa)/I_0(\kappa)\!=\!r$, where $I_0$ and $I_1$ were the modified
Bessel functions of zeroth and first order. The concentration parameter
$\kappa$ is a measure of phase locking; the phase distribution becomes uniform
for $\kappa\rightarrow 0$ and approaches a Gaussian distribution with variance
$\sigma^2\!=\!1/\kappa$ for large $\kappa$.

\subsection*{Information in oscillatory spike trains}

The estimated phase $\phi(t)$ of retinal oscillation at the time of stimulus
onset, $t$, was used to align the timing of thalamic spikes across all trials
by shifting each trial in time, $-\phi(t)/(2\pi f)$ (the absolute value was
$<10$~ms). Thus, the phase of the retinal oscillations, which had been
randomly distributed across trials, was made the same for each
one~\ref{fig:3}).  This alignment was made once for each cycle of the retinal
oscillation in order to ``de-jitter'' the entire thalamic spike train (see
Fig.~\ref{fig:3}D). Note that our method of de-jittering differs from those
that use the stimulus~\citep{Aldworth2005} or the spike train
itself~\citep{Richmond1990} as references in time. After aligning phases, we
used equation~(\ref{eq:inf-gauss}) to provide an estimate of information rate
that included the contribution of oscillations that were not locked to the
stimulus. This method has been verified previously~\citep{Koepsell2008a} by
expanding the direct method of estimating information rates for single spikes
to take the phase of oscillations into account.

\subsection*{Simulation Experiments}

Recently, \citep{Koepsell2008a} we designed the {\em quasi-periodic gamma
  (QPG) model} to understand how the two different information channels might
be multiplexed in the response of relay cells, (Fig.~\ref{fig:9}A): The QPG
model describes spike generation by an inhomogeneous Gamma
process~\citep{Barlow1957} with a factorial instantaneous rate. The
conditional probability of generating a new spike at time $t_i$, given the
last spike at time $t_{i-1}$ is written as~\citep{Barbieri2001}
$$
p_t(t_i|t_{i-1})=\frac{k\lambda(t_i)}{\Gamma(k)}
  \left[k\int_{t_{i-1}}^{t_i} \lambda(u) du\right]^{k-1}
  \exp\left\{-k\int_{t_{i-1}}^{t_i} \lambda(u) du\right\}\,,
$$
where $k$ is the shape parameter of the gamma distribution, $\Gamma(k)$ is the
gamma function, and the instantaneous rate $\lambda(t)$ is given by the
product
$$
\lambda(t) = 2\pi(RF \otimes s(t)) M(\phi(t); \kappa,\mu)\,.
$$
The first factor $RF\otimes s(t)$ is the averaged firing rate calculated by
convolving the stimulus with the receptive field,estimated as above). The
second factor is a von Mises distribution $M(\phi;\kappa,\mu)$ that describes
the periodic activity in single trials. The instantaneous phase of the
periodic activiy, $\phi(t)$, is given by the phase a random band-pass signal
with frequency $f\pm\sigma_f$. All told, in addition to the parameters
describing the receptive field, the model has four free parameters: $k$,
$\kappa$, $f$, $\sigma_f$. (An additional degree of freedom, the mean phase,
was arbitrarily set to $\mu\!=\!0$).

\subsubsection*{Fitting the model parameters}

To assess how well the results in Figs.~\ref{fig:1}-\ref{fig:8} were captured
by the QPG model, we fitted the free parameters of the model to match the
properties of the cell whose responses are indicated by the circled points in
Figs.~\ref{fig:8}A, B.  The concentration parameter $\kappa$ was determined by
fitting the von Mises distribution (Fig.~\ref{fig:7}B) to the phase
distribution of the cell's spikes and the parameters $f$ and $\sigma_f$ were
fitted to the spectrogram of the spikes (see section~\ref{sec:inf}). The shape
parameter k of the Gamma process was determined as follows. The averaged rate
$\lambda_0(t)$ was estimated from the average of the neural responses across
trials recordings by adaptive kernel estimation~\citep{Richmond1990}. After
rescaling time with
$$
t' = k\int_0^t \lambda_0(u) du
$$
in order to obtain a constant rate ($\lambda\!=\!1$), the rescaled
distribution of inter-spike intervals $\tau$ from the experimental data could
be fitted by a homogeneous Gamma distribution~\citep{Kuffler1957}
$$
p(\tau) = \frac{\lambda^k\tau^{k-1}e^{-\lambda\tau}}{\Gamma(k)}
$$
with fixed rate $\lambda\!=\!1$, shape parameter k and the gamma function
$\Gamma(k)$. The shape parameter was determined from the moments (mean and
variance) of the empirical rescaled distribution of the inter-spike
intervals~\citep{Barlow1957,Barbieri2001}
$$
k = \overline{\tau}^2/\sigma_\tau^2\,.
$$

\subsubsection*{Simulation results}

The fitted parameters for the cell encircled in Figs.~\ref{fig:8}A,B were
$k\!=\!2$, $\kappa\!=\!2.3$, $f\!=\!59$~Hz, $\sigma_f\!=\!2$~Hz and
$\mu\!=\!0$. The QPG model reproduced the multi-peaked distribution of
inter-spike intervals and the location of the tallest peak (near 17 ms) that
were observed empirically, compare Fig.~\ref{fig:9}B and Fig.~\ref{fig:2}A.

We then used the QPG model to estimate the rate at which information was
transmitted.  To reproduce the differences in phases among trials, we
randomized the phase offset $\phi_0$ for each one.  Again, the simulations
resembled the empirical results, compare Fig.~\ref{fig:10}A with
Fig.~\ref{fig:6} and Fig.~\ref{fig:10}C with Fig.~\ref{fig:6}C. Next, to
simulate information rates and spectra after de-jittering, we fixed the phase
offset $\phi_0$ in the model for all trials. Once more, the QPG model
reproduced the experimental results.  This similarity is seen in the shape of
the spectra and in the information rate (compare Fig.~\ref{fig:10}B with
Fig.~\ref{fig:6}B); the total upper bounds in bit per spike before (0.51,
model and 0.44, experiment) and after de-jittering (1.66, model and 1.62,
experiment).

Finally, we used the QPG model to generate surrogate datasets that comprised
many more trials than could be acquired with whole-cell recording in vivo.  We
applied the direct method to the simulated datasets to compare information
rates before and after de-jittering, Figs.~\ref{fig:10}C, D. This analysis
supported our conclusion that neural oscillations carry information
downstream.  That is, the information rate was 0.6 bit/spike (linear
extrapolation to zero bin, Fig.~\ref{fig:10}C dashed line) when the phase of
the oscillations wwas random across trials but grew to 1.8~bit/spike, when the
phase of the oscillations phase was aligned across trials, Fig.~\ref{fig:10}D,
dotted line.

\subsection*{Acknowledgments} 
We thank the former members of the Redwood Neuroscience Institute for
discussions and are grateful to Matthias Bethge, Tim Blanche, Yang Dan,
Bartlett Mel, Bruno Olshausen, Jascha Sohl-Dickstein, and David Warland for
comments on previous versions of the manuscript. This work was supported by
the National Institutes of Health (JAH, grant NIH EY09593), and the Redwoood
Neuroscience Institute (FTS). The clustering of intracellular events was
performed using the Klustakwik~\citep{Harris2000} program. The receptive field
estimation was done with MATLAB (TM) programming language. The remaining
analysis was performed using IPython~\citep{Perez2007} and
NumPy/SciPy~\citep{Oliphant2006,Oliphant2007}, an open-source software
environment written in Python. All figures were produced using
Matplotlib~\citep{Barrett2005}.

\subsubsection*{Author Contributions} KK conceived the idea for the project and
developed as well as conducted the main analyses of the data. XW, YW and VV
developed methods to detect events in the intracellular signal and, with JAH,
performed the experiments. QW and XW contributed software for stimulus
presentation, data collection and analysis. DLR and WMU performed the
extracellular experiments. FTS supervised the project at both sites and
contributed to both the theoretical analyses and the development of new
methods to analyze the data. The manuscript was written by KK, FTS and JAH.

\bibliographystyle{apalike} 

\end{document}